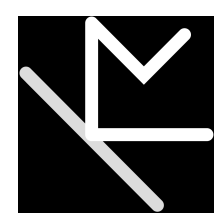

# JURISDICTION OVER UBIQUITOUS COPYRIGHT INFRINGEMENTS: SHOULD RIGHT-HOLDERS BE ALLOWED TO SUE AT HOME?

**Toshiyuki Kono ***

**Paulius Jurčys ****

## Abstract

The emergence of the Internet, and more recently cloud computing, has had significant technological, economic, social, and cultural effects. These developments affect the legal framework and call for careful assessment of whether, and how, existing legal principles and doctrines should be adjusted. Although cloud-based technologies now shape almost every aspect of communication, discussion of their legal implications remains fragmentary.

This paper focuses on the intersection between private international law and intellectual property rights in the cloud environment. Although the Internet is one of the most economically important markets for exploiting intellectual property, the ubiquity of the World Wide Web also creates legal risks. One such risk concerns litigation over the exploitation of intellectual property rights before foreign courts.

In private international law terms, this risk concerns international jurisdiction: in disputes between parties from different states, or disputes involving foreign subject matter, which court should hear the case? Under what conditions should a national court exercise jurisdiction over a multi-state dispute? National laws usually provide rules or principles to guide courts in making this determination, such as the defendant's residence or the commission of tortious acts in the forum state.

Jurisdiction in multi-state intellectual property disputes has long been controversial. Even leading courts have struggled with complex cases involving the cross-border exploitation of intellectual property rights. These difficulties have only intensified with the rise of digital communication technologies.

This chapter discusses these issues by first illustrating how cloud computing affects the exploitation of intellectual property assets. It then analyzes the main principles used by courts across the Atlantic to determine jurisdiction in multi-state intellectual property disputes. The fourth section asks whether the existing legal framework is suitable for cloud-related controversies. Finally, the chapter discusses the work of a special committee established under the auspices of the International Law Association.

* Professor of Law, Faculty of Law, Kyushu University (Japan); Vice-President of ICOMOS.

** LL.M. (Harvard), LL.D. (Kyushu). This article is based on a presentation made at the ALAI Congress in Kyoto (16-18 October 2011). The authors would like to thank Professors William Fisher, Christopher Bavitz, Helen Hershkoff, Shinto Teramoto, Benedetta Ubertazzi, Marcelo Corrales, as well as the Orin Foundation for help in preparing this article.

## Table of Contents

## 1. Migration to the Cloud

When the ALAI Congress was held in Kyoto in October 2012, Pirate Bay announced its decision to move to the cloud. This comes as no surprise as the Swedish courts[1] ordered to close down Pirate Bay in Sweden, a dozen EU countries, as well as the United States, China, India, and Malaysia. The Pirate Bay controversy raised questions about the technical feasibility of closing its websites, the desirability of balancing the interests of proprietors of IP rights with such values as freedom of speech and dissemination of information. Pirate Bay considered such legal actions "free advertising."[2] In a statement about its migration to the cloud and technical impossibility to shut down its service platform, Pirate Bay proclaimed that:

> *If the police decide to raid us again there are no servers to take, just a transit router. If they follow the trail to the next country and find the load balancer, there is just a disk-less server there. In case they find out where the cloud provider is, all they can get are encrypted disk-images*.[3]

The emergence of the Internet had tremendous technological, economic, social as well as cultural effects. More recently, cloud-based technologies have swiftly reached almost every aspect of communications. This paper focuses on a rather specific aspect concerning the intersection of private international law and intellectual property (IP) in the cloud environment. Although the Internet is one of the most economically rewarding markets for the commercialization of IP rights, the ubiquity of the World Wide Web is also associated with a number of risks. One of the risks that should be considered by right holders and online intermediaries concerns potential litigation over cross-border IP matters. In private international law terms, a number of questions arise: which court should adjudicate the case? Under what conditions can a court of one state exercise its jurisdiction and decide a multi-state dispute? Once the case is decided, could a court judgment be recognised and enforced abroad? The exercise of jurisdiction in multi-state

---

[1] *See* decisions of Stockholm's District Court (Tingsrätten) case B 13301-06, 17 April 2009 and Svea Court of Appeals (Hovrätten), case B 4041-09, November 26, 2010.

[2] See *Belgian ISP Ordered to Block the Pirate Bay*, activepolitic.com:82/News/2011-10-04a/Belgian_ISPs_Ordered_To_Block_The_Pirate_Bay.html.

[3] *Pirate Bay Moves to the Cloud, Becomes Raid Proof*, torrentfreak.com/pirate-bay-moves-to-the-cloud-becomes-raid-proof-121017. Pirate Bay was shut for few months after the Swedish Police raided a data center in Stockholm in December 2014, but the site – with a new logo of Phoenix – was re-launched again on January 31, 2015, *see* venturebeat.com/2015/01/31/the-pirate-bay-is-back-online-after-almost-two-months//.

IP disputes have been subject to vehement discussion.[4] Even the most distinguished courts in various countries stumbled when dealing with intricate quandaries involving cross-border exploitation of IP rights. One of the most notable examples arose when Apple and Samsung entered into litigation proceedings in three different continents.[5] Proceedings between Apple and Samsung once again brought to light the inefficiency of parallel proceedings in multi-state IP disputes.

The present paper begins with a short illustration on how cloud computing disrupted traditional business models involving IP rights and their protection. This discussion provides a solid background for a closer analysis of the main principles employed by the courts across the Atlantic in deciding when to assert jurisdiction over multi-state IP disputes. The paper highlights core doctrines such as subject-matter and exclusive jurisdiction over claims concerning foreign IP rights as well as various tests employed by the courts in order to assert long-arm jurisdiction over non-resident defendants. In light of the proliferation of cloud-based services, this paper examines the question whether, and if so, under what circumstances, a court of the state where the copyright holder resides could assert jurisdiction over non-resident defendants. It is suggested that concentrating litigation over copyright infringements before the right holder's home court could, in certain cases, provide a workable and well-balanced solution in determining jurisdiction over multi-state IP infringements.

## 2. Cloud Computing and its Effects on Creative Industries

### 2.1. New Technology, New Business Models

For the purposes of this article, "cloud computing" refers to a situation where a number of electronic devices such as computers, mobile phones and tablets are connected to a network. The information is stored on a remote server (the "cloud") from and through which the information is disseminated. A peculiar feature of cloud computing is that users of cloud services are often unaware that their data is stored in the cloud.

[4] *See*, *e.g.*, Toshiyuki Kono (ed.), Intellectual Property and Private International Law: Comparative Perspectives (2012); Paul Torremans (ed.), Research Handbook On Cross-Border Enforcement Of Intellectual Property (2015); European Max Planck Group on Conflicts of Laws in Intellectual Property, The CLIP Principles and Commentary (2013); Benedetta Ubertazzi, Exclusive Jurisdiction in Intellectual Property Law (2012); James Fawcett & P Torremans, Intellectual Property and Private International Law (2011); Jürgen Basedow, Toshiyuki Kono & Axel Metzger (eds.), Intellectual Property in the Global Arena (2010); Dario Moura Vicente, La propriété intellectuelle en droit international privé (2009); Stefan Leible & Ansgar Ohly (eds.), Intellectual Property and Private International Law (2009); Anette Kur & Josef Drexl (eds.), Intellectual Property and Private International Law (2005).

[5] As for the litigation in the U.S., some court documents in the dispute between Apple and Samsung can be found here: http://cand.uscourts.gov/lhk/applevsamsung.

Among digital communication technologies of the 1970s, only the large mainframe computers of universities and large corporations were connected to terminals. The information was stored in a central server that was physically located in the same place as the other components of the computer. At that time, data was processed in the central terminal and not in the mainframe computers. In the 1980s, personal computers with the ability to process data began to spread, and their processing and storage capacities increased rapidly. The development of the Internet in the 1990s became the key factor for incremental growth of communication speed as well as the capacities of communication bandwidth. Software developers flourished by bringing to market various data processing programs and applications. This was followed by the emergence of cloud computing, which marks a technological shift in which data storage servers are separated from the computer and the information can be synchronized between different devices with different users.[6]

Cloud computing could be understood as a method of providing various digital services through networks connected through the Internet. Perhaps one of the more precise definitions was offered by the US National Institute of Standards and Technology (NIST), which describes cloud computing as "a model for enabling convenient, on-demand network access to a shared pool of configurable computing resources (e.g., networks, servers, storage, applications, and services) that can be rapidly provisioned and released with minimal management effort or service provider interaction." [7] Nonetheless, there is currently no single uniformly accepted definition of cloud computing. Instead, similar definitions such as grid computing, utility computing, or ubiquitous computing are often used interchangeably.

Functionally, it is possible to consider cloud computing as a technological improvement that opens possibilities for the development of a number of new business models. These new kinds of business models make use of cloud computing technology to offer various kinds of services: software as a service (SaaS); platform as a service (PaaS); infrastructure as a service (IaaS), storage as a service (StaaS), data as a service (DaaS); database as a service (DBaaS); or security as a service (SECaaS). A common feature of such services is that the information is stored on a remote server. In practice, a number of these cloud-based services are provided in a bundle. For instance, users of blogging services usually use software, storage, and a platform provided by an online intermediary.

---

[6] SHINTO TERAMOTO (ED.), KURAUDO JIDAI NO HŌRITSU JITSUMU [LEGAL PRACTICE IN THE CLOUD ERA] 3-4 (2010).

[7] See csrc.nist.gov/publications/nistpubs/800-145/SP800-145.pdf.

Even in the absence of a common definition, it is possible to distinguish the following characteristics of cloud computing: (a) high scalability (i.e., ability of a network or a system to handle a growing amount of work in a capable manner or its ability to be enlarged to accommodate that growth); (b) abstracted computer resources (i.e., the possibility to geographically separate a personal data processing device from the information storage device); (c) the possibility of being provided as a service; (d) low costs of exploitation.[8] The reduction of data processing costs has remarkable market effects. In the context of creative and technological industries, cloud computing facilitates the distribution of larger amounts of IP-protected goods to a broader range of users and at a lower price.

## 2.2. Exploitation of IP Rights and Associated Risks in the Cloud

International trade rests upon the existence of differences among national economies. The possibility to exchange capital for resources is in the interests of trading nations, which can then focus their economic activities to achieve economies of scale.[9] During the twentieth century, economic ties between nations became even more closely knit, as the process of globalization accelerated when digital communication technologies became widely accessible. The costs associated with the movement of information, capital, and persons have been curtailed to a significant degree by the introduction of the Internet and cloud computing. This has collateral effects on national economies, cultural and social life.

IP assets often play a vital role in cloud-based global business models. For example, iTunes is an online platform where copyright-protected artistic works as well as software are distributed to subscribers for a certain price or even for free. Another example could be blogging. An online intermediary (such as Wordpress or Google) offers software that allows the user to create a personal blog and make it publicly available under the intermediary's subdomain or under the personalised domain. The user of the blogging services is then able to upload and publish posts which could contain links to other sites as well as upload media files. In both cases, the information is stored on servers in the cloud. There are also cloud-based business models offering platforms for the exchange of information. The users of such platforms, such as flickr.com, offer online space for sharing information or digital media files. There are also platforms, such as eBay and Amazon, that facilitate the exchange of specified assets. Social networking sites such as Facebook, LinkedIn, and Instagram also function as platforms for the exchange of information between the users. In most of these cases, the exchange of information occurs in the cloud.

---

[8] TERAMOTO, *supra* note 6, at 4-5; and Seiichirō Sakurai, *Ubikitasu shingai to kuraudo konpiūtingu [Ubiquitous Infringement and Cloud Computing]*, *in* CHITEKI ZAISAN-KEN TO SHŌGAI MINJI SOSHŌ [INTELLECTUAL PROPERTY RIGHTS AND INTERNATIONAL CIVIL LITIGATION] 374 (Toshiyuki Kono ed., 2010).

[9] PAUL R. KRUGMAN, MAURICE OBSTFELD & MARC J. MELITZ, INTERNATIONAL ECONOMICS: THEORY AND PRACTICE 24 (2012).

The Internet and cloud computing also had tremendous effects on the publication business. In conventional distribution environment where the printed material or records are published in hard copies, the creators usually have to contact the publisher or a record company who mainly obtain control over publishing and distribution of the printed material. This process of publication is usually lengthy and rather expensive. Moreover, the creator has fairly limited possibilities to affect the distribution of his works. The process of publishing looks rather different in the Internet and cloud-based environment. In the case of digital publishing, the right holder has much more possibilities to control the modes of publication of his or her works by choosing the preferred platform(s) where the work should be published (e.g., soundcloud.com). These platforms are offered by online intermediaries, but one of the main differences from conventional mode of publication is that digital publishing allows the creator to publish the work instantaneously, at virtually no cost, and reach the broader audiences.

The exploitation of IP in the cloud computing environment in all cases involves the interests of at least three stakeholders: the originator of information (IP right holder[10]), users of information, and the intermediary who provides cloud-based services.[11] From a private international law perspective, the aggravating factor is the fact that the parties to a dispute could be from different countries. Assume that the originator of information (actor a) is resident in country A whereas the user (actor b) is resident in country B. The communication between actor a and actor b happens via intermediary (actor c) who may be resident in a third country. Yet, even if all three stakeholders were resident in the same country, the question would be whether the location of servers in the cloud affects the functioning of private international law rules allocating jurisdiction and determining the governing law.[12]

[10] The notion of a 'right holder' in the cloud could be used also in a broader context and cover not IP right holders, but any other person who originates information (*e.g.*, author, singer, or a doctor who saves information about the patient in the database of national hospitals).

[11] Marc A. Melzer, *Copyright Enforcement in the Cloud*, 21 FORDHAM INTELL. PROP. MEDIA & ENT. L.J. 403 (2011); Marcelo Corrales, *Databases in the Cloud: A Sui Generis Contractual Model for Cloud Brokerage Scenarios*, papers.ssrn.com/sol3/papers.cfm?abstract_id=2410740.

[12] *See*, *e.g.*, Damon C. Andrews & John M. Newmann, *Personal Jurisdiction and Choice of Law in the Cloud*, 73 MD. L. REV. 313 (2013).

The notion of cloud computing could be understood as referring to the technical architecture of interconnected networks; while ubiquity of a copyright infringement implies that certain activities could lead to alleged infringements in every state in which the content is available. From a legal point of view, a number of conflict of laws questions arise. For example, which law should be applicable to the activities of the actors? Does the fact that the communication between different actors is transmitted through the servers in the cloud alter the determination of law applicable to the activities and relationships between the users of cloud-based services? In the same vein, similar jurisdiction-related questions could be posed: on what grounds should the courts be able to assert jurisdiction over cloud-related disputes? More generally, what are the legal implications of the fact that the information is stored in the cloud? It is possible to distinguish three different types of controversies:

a) *Disputes between the IP right holder and the alleged infringer*. In such cases the IP right holder usually seeks legal redress against the user for an alleged infringement. For example, a copyright holder might sue a third party for an unauthorised uploading and sharing of digital copies of a work, claiming the infringement of the right to make the work publicly available. The court then has to decide whether it is competent to adjudicate the case. But the difficulty arising in such a situation is usually related to the establishment of the necessary connection between the alleged infringement of copyrights and the forum (e.g., the residence of the copyright holder in the forum state, or availability of the infringing content in the forum state);

b) *Disputes between the IP right holder and cloud service provider (intermediary)*. The territorial nature of IP rights makes it very difficult for an IP right holder to efficiently enforce IP rights. This ability to protect rights is even more aggravated by cloud-based services which can be dynamically relocated at any point in time. Due to such technological challenges, an IP right holder may find it more practical to sue the intermediary whose cloud services are used to facilitate IP infringements. From the efficiency perspective,[13] the possibility of suing the cloud service provider is an attractive option:[14] the IP right holder can file a single action before the court of the intermediary's habitual residence or in a state where the intermediary conducts the majority of its commercial / infringing activities;

[13] Shinto Teramoto, *"Nihon-hō no tōmeika" purojekuto ni yoru rippō teian e no komento: CLIP gensoku dai ni jian oyobi ALI gensoku to no hikaku ni oite'* ['Legislative Proposal of the "Transparency of Japanese Law" Project: Comments Based on the Second Draft Proposal of the CLIP Principles and the ALI Principles'] *in* KONO (ed.), CHITEKI ZAISAN-KEN TO SHŌGAI MINJI SOSHŌ [INTELLECTUAL PROPERTY RIGHTS AND INTERNATIONAL CIVIL LITIGATION] 413 (2010).

[14] In this regard, *see* Graeme Dinwoodie, Rochelle C. Dreyfuss & Anette Kur, *The Law Applicable to Secondary Liability in Intellectual Property Cases*, (2009) 42 N.Y.U. J. INT'L L. & POL. 201 (2009); Pedro A. De Miguel Asensio, *Internet Intermediaries and the Applicable law to Intellectual Property Infringements*, 3(3) JIPITEC 350 (2013); Shinto Teramoto & Paulius Jurčys, *Intermediaries, Trust and Efficiency of Communication: A Social Network Perspective*, *in* NETWORKED GOVERNANCE, TRANSNATIONAL BUSINESS AND THE LAW 99 (Mark Fenwick, Steven van Uytsel & Stefan Wrbka eds., 2014).

c) *Disputes between service users and the provider of services in the cloud*. In such situations the parties are usually bound by contractual agreements and the subject matter of the dispute may concern the quality of services. It is quite likely that jurisdiction will be determined by the parties' ex ante choice of court agreement.[15] However, in the absence of a choice of court clause, the court will have to determine whether it has jurisdiction under the forum law.

From the three above-mentioned categories, most of the reported cases concern IP-right infringements. In the early years of the Internet, numerous cases concerned disputes between proprietors of IP rights and direct infringers. But due to various practical and legal considerations, an increasing number of cases are filed against intermediaries.[16] Despite the fact that such online IP infringements are very frequent, even the most distinguished courts from various countries have been struggling with the grounds on which jurisdiction could be asserted. As demonstrated in most of the national reports prepared for the 2012 ALAI Congress in Kyoto, one should not be astonished that there are no special jurisdiction and choice of law rules for copyright issues related to the cloud environment. [17] Instead, it appears that general private international law rules established in national, regional or international legal instruments, if any have to be applied to cloud-related copyright issues.[18]

[15] *See* Art. 5(1) of the 2005 Hague Convention on Choice of Court Agreements, *available at*: hcch.net/upload/conventions/txt37en.pdf.

[16] *See*, *e.g*., A&M Records, Inc. v. Napster, Inc., 239 F.3d 1004 (2001); In re Aimster Copyright Litigation, 334 F.3d 643 (7th Cir. 2003); MGM Studios, Inc. v. Grokster, Ltd., 545 U.S. 913 (2005).

[17] *See Dutch Report*, alai.jp/ALAI2012/program/national_report/Netherland.pdf, at 16; *Finnish Report*, alai.jp/ALAI2012/program/national_report/Finland.pdf, at 9-10; *German Report*, alai.jp/ALAI2012/program/national_report/Germany.pdf, at 48-58; *Mexican Report*, alai.jp/ALAI2012/program/national_report/Mexico.pdf, at 10; *Spanish Report*, alai.jp/ALAI2012/program/national_report/Spain.pdf, at 18; *UK Report*, alai.jp/ALAI2012/program/national_report/Netherland.pdf, at 11-12.

[18] *See Belgian Report*, alai.jp/ALAI2012/program/national_report/Belgium.pdf, at 50-56; *Croatian Report*, alai.jp/ALAI2012/program/national_report/Croatia.pdf, at 13-15; *French Report*, alai.jp/ALAI2012/program/national_report/France.pdf, at 13-16; *Greek Report*, alai.jp/ALAI2012/program/national_report/Greece.pdf, at 7; *Italian Report*, alai.jp/ALAI2012/program/national_report/Italy.pdf, at 11; Norwegian Report, alai.jp/ALAI2012/program/national_report/Norway.pdf, at 7; *Polish Report*, alai.jp/ALAI2012/program/national_report/Poland.pdf, at 7-8; *Portuguese Report*, alai.jp/ALAI2012/program/national_report/Portugal.pdf, at 13; *Swedish Report*, alai.jp/ALAI2012/program/national_report/Sweden.pdf, at 8; *Swiss Report*, alai.jp/ALAI2012/program/national_report/Switzerland.pdf, at 8-9; *US Report*, alai.jp/ALAI2012/program/national_report/UnitedStates.pdf, at 36-16.

## 3. Jurisdiction over Cross-border Copyright Infringements

Difficult questions of jurisdiction arise in cases involving parties from different countries, or where alleged acts are committed in multiple places or have cross-border effects. Due to the ubiquity of the Internet, the flow of information stored in the cloud becomes even more borderless. However, the emergence of global business models has had little impact on the legal notions of sovereignty and territorial reach of national laws. In order to assert jurisdiction over a case, courts still need to find a territorial connecting factor (e.g., residence of parties, location of property or the place where certain acts were committed). Such territorial connecting factors are considered as a signal justifying the existence of a close connection between the substance of the dispute and the forum state.

The following sections illustrate peculiar features in common law and civil law countries concerning the exercise of jurisdiction in multi-state IP disputes. In particular, two constitutive elements of establishing jurisdiction over copyright infringements in common law and civil law are compared: (i) jurisdiction over the subject matter; and (ii) jurisdiction over persons. This comparative analysis helps delineate main considerations that common law courts and civil law courts employ in determining whether jurisdiction should be asserted. Comparative findings will bring into view some of the recent trends in litigation with regard to cross-border copyright infringements and set the stage for the discussion in the subsequent section on whether there should be some relaxation in the traditional approach by opening some possibilities for the copyright holder to seek cross-border relief at the home court.

### 3.1. Jurisdiction over the Subject Matter

The first issue is related to the so-called "subject-matter" or "exclusive" jurisdiction. Historically, printing and publishing businesses were based on exclusive privileges granted by kings and princes. Such privileges were granted to certain guilds on an individual basis, often for some kind of pecuniary consideration. Thus, kings were able to determine what was printed and by whom. Naturally, such exclusive rights were effective only within the territories of sovereigns who were granting such rights.[19] Later, as nation states started to adopt specific IP statutes, exclusive printing and publishing privileges lost their significance. A system of IP rights gradually developed in a more institutionalised framework where *ex ante* requirements to obtain protection were entrenched in the statutes. The fact that IP rights were granted on the basis of national law meant that such rights were legally effective only within the territory of the granting state.

[19] *See* Jürgen Basedow, *Foundations of Private International Law in Intellectual Property*, *in* BASEDOW, KONO & METZGER (EDS.), *supra* note 4, at 3-29.

The existing legal framework has been harmonised by a number of multilateral treaties, most of which form the so-called TRIPs Plus system.[20] Although several treaties have been adopted with the objective of harmonising the standards of the protection of copyright,[21] none of them contain jurisdiction provisions.[22] The only related rule could perhaps be Article 5(2) of the Berne Convention, which provides inter alia that "the extent of protection, as well as the means of redress afforded to the author to protect his rights, shall be governed exclusively by the laws of the country where protection is claimed."

Much attention has been devoted to the actual purpose and meaning of this provision.[23] Here, it suffices to note that there is no agreement as to whether Article 5(2) provides for any guidance in terms of determining which country's court should hear the case. The principle of national treatment in Article 5(1) of the Berne Convention appears to be more important, since it requires member states to make sure that foreign authors are able to enjoy the same rights as the national authors do. This requirement to treat foreign and national authors equally eliminates the "foreignness" of a foreign right holder. Thus, in multi-state copyright disputes the other "foreign" element is one or more foreign copyrights for which the author seeks protection.

The following sections are devoted to providing a closer exposition of the approaches taken by the courts in common law countries and civil law countries. In common law, courts have for a long time considered claims related to foreign IP rights as not justiciable.[24] For a number of reasons common law courts usually decide that they do not possess subject-matter jurisdiction over claims concerning foreign IP rights. In civil law countries, notions such as justiciability or subject-matter jurisdiction are unfamiliar. Nevertheless, procedural statutes contain rules conferring exclusive jurisdiction upon certain courts and which parties cannot escape by making a choice of court agreement. Exclusive jurisdiction rules also override general jurisdiction at the defendant's forum.

---

[20] *See*, *e.g.*, DANIEL GERVAIS, THE TRIPS AGREEMENT (2012); CARLOS CORREA, TRADE RELATED ASPECTS OF INTELLECTUAL PROPERTY RIGHTS (2006).

[21] Most notably, 1886 Berne Convention for the Protection of Literary and Artistic Works; 1961 Rome Convention for the Protection of Performers, Producers of Phonograms and Broadcasting Organizations; 1996 WIPO Copyright Treaty (WCT); and 1996 WIPO Performances and Phonograms Treaty (WPPT).

[22] *See*, *e.g.*, SAMUEL RICKETSON & JANE C. GINSBURG, INTERNATIONAL COPYRIGHT AND NEIGHBOURING RIGHTS (2005).

[23] For a closer overview of the discussion *see* Toshiyuki Kono & Paulius Jurčys, *General Report*, *in* INTELLECTUAL PROPERTY AND PRIVATE INTERNATIONAL LAW: COMPARATIVE PERSPECTIVES 6 (Toshiyuki Kono ed., 2012).

[24] *See*, *e.g.*, Richard Fentiman, *Choice of Law and Intellectual Property*, *in* DREXL & KUR, *supra* note 4, at 129-50; Kurt Lipstein, *Intellectual Property: Jurisdiction or Choice of Law?*, 61(2) C.L.J 300 (2002).

### *(a) Jurisdiction over the Subject Matter in Common Law Countries*

In order to exercise adjudicative authority over a dispute, a common law court must establish both the existence of subject-matter jurisdiction as well as personal (in personam) jurisdiction. In personam jurisdiction requires that a court determine whether the defendant has sufficient contacts with the forum state to justify the exercise of the court's authority over that defendant. Subject-matter jurisdiction requires that a court determine whether it has jurisdiction over the substance of the case.[25] In practical terms, the requirement of subject-matter jurisdiction means that common law courts are not competent to hear disputes concerning the validity and infringement of foreign IP rights. As a result of such a strict approach, the adjudication of multi-state IP disputes must take place before courts of every state for which the protection is sought.

The common law approach toward the adjudication of disputes involving foreign IP rights could be best explained from a historical perspective. In some early multi-state patent cases,[26] common law courts drew parallels between patent infringement claims and claims related to foreign land. The granting of a patent right was considered an "act of state."[27] Accordingly, infringement actions were considered justiciable only before the courts of the granting state. The notion of non-justiciability of foreign IP-related claims was echoed in the practice of other common law countries not only in patent cases,[28] but also in disputes pertaining to other kinds of IP rights.

In *Subafilm*, a US copyright infringement case, the dispute arose with regard to the distribution of an animated motion picture titled *Yellow Submarine* which also contained the famous Beatles single. Pursuant to initial agreements, United Artists Corporation obtained rights to distribute the film in theaters as well as television. After the emergence of videocassettes, the successor company of the licensee started to distribute copies of the *Yellow Submarine* internationally, also in the form of videocassettes. Because the initial contracts between the parties did not contain any clear provisions regarding videocassettes, the plaintiffs filed an action seeking monetary relief for loss sustained in the United States and abroad from selling the videos for home use. The US Court of Appeals followed its established practice and held that it would not hear claims related to the infringement of copyrights in foreign states. In particular, the Court noted that US copyright laws do not extend to acts that occur outside of the United States and that the copyright holder is able to recover only "damages that stem from a direct infringement of its exclusive rights that occurs within the US."[29]

---

[25] Ralphe A. Armstrong and Anna Music v. Virgin Records Ltd., 91 F. Supp. 2d 628, 636 (2000). For the application of these jurisdiction requirements in patent infringement disputes *see* Viam Corp. v. Iowa Export-import Trading Co., 84 F.3d 424, 427 (Fed. Cir. 1996).

[26] Potter v. Broken Hill Pty. Co. Ltd., V.L.R. 612 (1905), *aff'd* 3 C.L.R. 479 (1906).

[27] *Id.* at 494.

[28] As for the UK, *see*, *e.g*., Tyburn Productions Ltd. v. Conan Doyle, 19 I.P.R. 455 (1990); Coin Controls Ltd. v. Suzo International (UK) Ltd., 3 All E.R. 45, 52 (1997).

The strict territorial approach for adjudicating multi-state copyright disputes was often a subject of controversy. Though mosaic state-by-state litigation was criticized for its high costs, common law courts have not been keen to change their long-standing jurisprudence.[30] The recent *Lucasfilm* case decided by the UK Supreme Court in 2011 offers some hope for advocates of a more flexible approach toward adjudication claims concerning foreign IP rights and possibilities of consolidation. At the center of this dispute was the protection of some of the Imperial Stormtrooper helmets that were used in *Star Wars* – a 1977 movie which won an Academy Award for Best Costume Design. George Lucas, the creator of the concept of Imperial Stormtroopers, brought an action against Andrew Ainsworth in the UK. Ainsworth used his original tools (which were the same used in making the actual costumes for *Star Wars*) to make Imperial Stormtrooper helmets and armor for sale to the public. The UK Supreme Court had to address the question of whether the claim against a defendant domiciled in England, for infringement of a foreign copyright, was justiciable or not.

Having reviewed a number of previous judgments, the Supreme Court came to the conclusion that there were no impediments for English courts to hear actions for the infringement of foreign IP rights.[31] In the same vein, the Court also ruled that the act of state doctrine would no longer be applicable in adjudicating infringement actions under foreign copyright statutes [32] . Furthermore, the recent developments regarding the application of Article 24(4) of the Brussels I Regulation (Recast), the lex loci protectionis rule entrenched in Article 8 of the Rome II Regulation as well as the approach adopted in the ALI Principles,[33] were inferred as not forestalling the Court to hear claims related to the infringement of the US copyright.[34] In light of such considerations, the UK Supreme Court came to a conclusion that a claim concerning the infringement of a foreign copyright is justiciable provided that in personam jurisdiction exists.[35] This Supreme Court judgment has instigated many discussions, and some commentators anticipate that the courts of other common law countries will also follow the softer approach to justiciability of foreign IP rights.[36]

[29] Subafilms Ltd. v. MGM-Pathe Communications Co., 24 F.3d 1088 (9th Cir. 1994).

[30] *See* Jan K. Voda v. Cordis Corp. 476 F.3d 887 (Fed. Cir. 2007), where the plaintiff who was a proprietor of patents in multiple states in the US and Europe sought to consolidate claims before the US court; however, the Federal Circuit Court decided that considering principles underlying comity, judicial economy, convenience, fairness, independence of national patents and other exceptional circumstances there were no compelling reasons to adjudicate foreign patent infringement claims.

[31] *Id.*, at § 53-80.

[32] *Id*., at § 81-86.

[33] The American Law Institute, Intellectual Property: Principles Governing Jurisdiction, Choice of Law and Judgments in Transnational Disputes (2008), S. 301 *et seq*.

### *(b) Exclusive Jurisdiction in Civil Law Countries*

The general rule in civil law countries is that a claim against a foreign defendant should be launched before the courts of her domicile (residence). This general ground of jurisdiction however is not without exception. One of the exceptions usually concerns certain matters which are considered to be closely connected to one particular state (e.g., location of an immovable). In the context of IP, many civil law countries' procedural laws contain a provision which stipulates that claims related to validity or registration of IP should be brought before the courts of the country of registration.[37] In the EU, Article 24(4) positing exclusive jurisdiction rules for registered IP has been subject to much criticism especially after the CJEU's decision in *GAT v. LuK* case,[38] where the court held that exclusive jurisdiction rules also apply to situations concerning the validity of registered IP rights challenged by a defendant by way of counter-claim. By way of comparison, the Tokyo District Court in the "Coral Sand" case[39] held that the decision on invalidity of a foreign patent could only have limited effects among the parties.[40]

[34] *Id.* at § 87-94.

[35] *Id.* at § 105.

[36] *See* Joost Blom, *Canada*, *in* KONO, *supra* note 4, at 424-76; CHRISTOPHER WADLOW, *United Kingdom*, KONO, *supra* note 4, at 1061-1102; Paul Torremans, *The Sense or Nonsense of Subject Matter Jurisdiction over Foreign Copyright*, 33 E.I.P.R 349 (2011); Paul Torremans, *Star Wars Rids Us of Subject-Matter Jurisdiction: The Supreme Court does not like Kafka either when it Comes to Copyright*, 33 E.I.P.R. 813 (2011); Benedetta Ubertazzi, *Intellectual Property Rights and Exclusive (Subject-Matter) Jurisdiction*, GRUR 199 (2011); Paul Torremans, *Lucasfilm v. Ainsworth*, IIC 751 (2010); Graeme W. Austin, *The Concept of "Justiciability" in Foreign Copyright Infringement Cases*, IIC 393 (2009).

[37] *See*, *e.g.*, Art 24(4) of the Brussels I Regulation (Recast) which refers to 'proceedings concerned with the registration or validity of patents, trade marks, designs, or other similar rights required to be deposited or registered'; Art. 3-5(iii) of the Japanese Code of Civil Procedure (2012) refers to the "actions related to existence and effects" of IP rights registered in Japan.

[38] Case C-4/03, Gesellschaft für Antriebstechnik mbH & Co. KG (GAT) v. Lamellen und Kupplungsbau Beteiligungs KG (LuK), 2006 E.C.R. I-6509; *see also* Case C-616/10, Solvay SA v. Honeywell Fluorine Products Europe BV et. al., E.C.R. . Marketa Trimble, *GAT, Solvay and the Centralization of Patent Litigation in Europe*, 26 EMORY INT'L L. REV. 515 (2012).

The absence of the registration requirement for the existence of copyrights is one of the reasons why civil law countries' courts have usually asserted jurisdiction over claims involving foreign copyrights. This means that the copyright holders are able to bring claims not only before the defendant's forum, but also before the courts where copyright infringing acts were committed or where the harm was sustained. Such claims can be brought according to the special jurisdiction rules for infringement matters.[41] Nevertheless, this does not mean that the exercise of jurisdiction over disputes involving multi-state copyright infringements has been unproblematic. A number of intricate questions linger with regard to the scope of the court's competence in multi-state copyright infringement disputes, coordination of parallel proceedings as well as possibilities of joining the claims. The exercise of jurisdiction over disputes concerning multi-state/ubiquitous infringement of copyrights poses an array of additional problems related to the connecting factors based upon which the court can decide to hear a case.[42] Even if a court asserted jurisdiction over such dispute where the right holder seeks redress for the infringements of copyrights in multiple states, it would have to face a problem applying the law of multiple states. For instance, Art. 8 of the Rome II Regulation requires the application of the law of each state for which protection is sought.[43]

## 3. 2. Personal Jurisdiction over Copyright Infringements

The second jurisdictional issue related to disputes involving copyright violations in the cloud concerns jurisdiction over the parties. This issue has been one of the disputed points during the negotiations of the so-called Hague Judgments Project, which was undertaken by the Hague Conference on Private International Law.[44] The idea of drafting an international treaty which would harmonise certain aspects related to adjudication of international civil and commercial disputes and recognition of foreign judgments was proposed by the United States in the early 1990s. The drafting process took more than a decade, however, and the negotiating parties did not manage to find consensus. Among the key reasons for disagreement were certain legal concepts which had strong roots in national legal systems, such as doing business jurisdiction, domicile, and territoriality.[45] The negotiations ended in 2005 with the adoption of the Hague Convention on Choice of Court Agreements because this was perhaps the only issue upon which the negotiating parties could agree. Notably, in 2012, the Hague Conference announced reopening the Judgments project, stating that a number of jurisdiction-related issues which arose at the turn of millennium have been to a large extent clarified.

---

[39] Tōkyō Chihō Saibansho [Tōkyō Dist. Ct.] October 16, 2003, 1847 HANREI JIHŌ [HANJI] 23 (Japan); an abbreviated English translation is *available at* tinyurl.com/c9lsekr.

[40] For a more detailed discussion *see* Kono & Jurčys, *supra* note 23.

[41] Art. 7(3) of the Brussels I Regulation (Recast); Art. 3-3(viii) of the Japanese Code of Civil Procedure (2012) provides that "Japanese courts shall have jurisdiction over actions concerning unlawful acts if such acts occurred in Japan. However, this rule shall not apply with regard to infringing acts undertaken abroad, the effects of which occurred in Japan, if it could not have been generally foreseen that the effects of such acts will occur in Japan."

[42] Some of these issues will be addressed in the following sections of this paper.

[43] Regulation (EC) No. 864/2007 of the European Parliament and of the Council of 11 July 2007 on the law applicable to non-contractual obligations (Rome II), 2007 O.J. (L199) 40.

[44] For interim reports on various topics which were prepared during the negotiation of the Hague Judgments Convention see hcch.net/index_en.php?act=conventions.publications&dtid=35&cid=98.

The following section provides an overview of the differing approaches to personal jurisdiction in civil and common law countries. The analysis mainly focuses on possible jurisdictional questions which could arise in cases concerning the violation of copyright-protected works in the cloud environment. In particular, this section deals with the question of determining the relevant factors that could serve as grounds for exercising a court's adjudicative authority over copyright infringements occurring in the cloud. Until recently, there has surprisingly been no agreement in civil law countries as to the determination of the place of tort or injury for the purposes of asserting jurisdiction over the dispute. Even though some landmark judgments have been rendered in defamation cases, it remains questionable whether these approaches could be applied to copyrights and other IP rights. In order to facilitate the discussion, this section introduces some of the recent cases rendered by US Courts in copyright infringement cases, most notably the recent set of cases between Penguin and American Buddha.

### *(a) Personal Jurisdiction in Civil Law Countries*

In civil law countries, the general principle is that the action must be brought before the defendant's forum.[46] The general understanding is that conferring jurisdiction upon the courts of a defendant's domicile or residence makes it easier for the defendant to defend himself.[47] However, this general ground does not apply if there is a valid choice of court agreement between the parties designating that the courts of another state shall hear the dispute.[48] Defendant's forum rule also does not apply if the dispute should be adjudicated pursuant to exclusive jurisdiction rules. In civil law countries, certain kinds of disputes can usually be adjudicated before courts which have so-called special jurisdiction over the case, even if such courts are outside the defendant's forum. For instance, in infringement disputes, infringing acts or the harm caused by these acts could occur in a state or states where the defendant is not resident. Accordingly, courts in various civil law countries have held that courts in other states are better placed to adjudicate the dispute due to the location of evidence or witnesses.[49]

---

[45] *See* RONALD A. BRAND & SCOTT R. JABLONSKI, FORUM NON CONVENIENS: HISTORY, GLOBAL PRACTICE, AND FUTURE UNDER THE HAGUE CONVENTION ON CHOICE OF COURT AGREEMENTS (2007); FAUSTO POCAR & CONSTANZA HONORATI (ED.), THE HAGUE PRELIMINARY DRAFT CONVENTION ON JURISDICTION AND JUDGMENTS (2005); Rochelle C. Dreyfuss, *An Alert to the Intellectual Property Bar: The Hague Judgments Convention*, 1 U. ILL. L. REV. 101 (2001).

[46] Art. 4 of the Brussels I Regulation (Recast); Art. 3-2 of the Japanese Code of Civil Procedure (2012).

Most civil law countries do not have jurisdictional rules specifically tailored for multi-state copyright infringement disputes. Instead, general jurisdiction rules have to be applied. Often such rules are quite laconic. For instance, Article 7(3) of the Brussels I Regulation (Recast) stipulates that a "defendant who is resident in a Member State may also be sued before the courts of the place where the harmful event occurred or may occur." Similar provisions could be found in other civil law countries.[50] The practice of applying such rules to multi-state torts has developed over time. Early cases concerning environmental pollution raised the issue of identifying the location of the "harmful event" where the claim for damages could be brought. In *Bier*, CJEU indicated that the "place where the harmful event occurred" should be understood as covering both the place where the damage occurred, as well as the place of the event giving rise to damage.[51]

Problems related to the identification of the place of the harmful event or the harm itself in environmental cases are similar to copyright infringements in the cloud. If one looks to the situation in the European Union, it becomes clear that the practice of national courts differs significantly. Some thought-provoking considerations could be derived from earlier CJEU jurisprudence in cases concerning infringements of personality rights. In one of the milestone cases, *Shevill*,[52] a French newspaper published an article about a British national who was engaged in money laundering activities in France. The British national wanted to sue the French publisher in the United Kingdom on the basis of Article 7(3) seeking compensation for damages caused by the distribution of the newspaper in several other states. The CJEU decided that in order to be able to recover damages caused by the publication in multiple States, the plaintiff must file an action before the courts of the place where the defendant is domiciled or the place where the publisher is established. All other courts of the states where the publication was distributed (pursuant to Art. 7(3)) could hear only territorially limited claims concerning damage sustained in that particular state.

---

[47] *See*, *e.g.*, Case C-26/91, Handte v. Traitements Mécano-chimiques des Surfaces, 1992 E.C.R. I-3967, § 14.

[48] Art. 23 of the Brussels I Regulation (Recast); Art. 3-7 of the Japanese Code of Civil Procedure (2012).

[49] *See*, *e.g.*, Case C-220/88, Dumez France and Tracoba, 1990 E.C.R. I-49, § 17; or Case C-364/93, Marinari v. Lloyds Bank and Another, 1995 E.C.R. I-2719, § 10.

[50] Art. 3-3(viii) of the Japanese Code of Civil Procedure (2012) posits that: "Japanese courts shall have jurisdiction over actions concerning unlawful acts if such acts occurred in Japan. However, this rule shall not apply with regard to infringing acts undertaken abroad, the effects of which occurred in Japan, if it could not have been generally foreseen that the effects of such acts will occur in Japan."

[51] Case 21/76, Handelskwekerij GJ Bier BV v. Mines de potasse d'Alsace SA, 1976 E.C.R. 1735, §§ 24-25; Saikō Saibansho [Sup. Ct.], June 8, 2001, 55 SAIKŌ SAIBANSHO MINJI HAREISHŪ (MINSHŪ) 727 (Japan). The CJEU was reluctant to extend the notion of the "harmful event" so as to cover indirect damages. *See e.g.*, Case C-220/88, Dumez France SA and Traboca SARL v. Hessische Landesbank and others, 1990 E.C.R. I-49, § 22; Case C-364/93, Antonio Marinari v. Lloyds Bank plc and Zubaidi Trading Company, 1995 E.C.R. I-2709, §§ 14–15.

A more recent case, *eDate*, concerned the determination of the place of harm when content is distributed online. In *eDate*,[53] a French actor, Martinez, brought an action in Paris against a British media company seeking compensation for damages to his personality rights by publication of a news item stating that Martinez renewed his friendship with another famous Australian singer Kylie Minogue. The court addressed the issue of whether a victim of a defamatory publication online could bring an action before the court of his own habitual residence and, if so, what the scope of the court's powers would be in deciding the question of compensation for multistate damages. Referring to previous case law, the CJEU held that for the purposes of establishing jurisdiction, a distinction should be made between the regional distribution of printed material and online ubiquitous distribution. The Court held that infringements that occur on the Internet cause serious harm to the right holder because the information may be available on a worldwide basis.[54] Article 7(3) of the Brussels I Regulation (Recast) enables the plaintiff to bring an action in any court where the content was accessible, which is tantamount to the place where damage was sustained. However, in the absence of any other connecting factors, such a court will be able to decide only with regard to damages sustained in that forum state. A claim for recovery of all of the damage sustained could be brought before a court of the defendant's domicile (Art. 4) as well as the court of the state where the injured person "has his center of interests". The practical significance of *eDate* is that the Court added an additional ground of jurisdiction where the victim could recover the damage sustained in multiple states. In order to justify this kind of policy decision, the CJEU referred to the need to assure efficacious conduct of the proceedings[55], objective and sound administration of justice[56], predictability[57], as well as "the existence of a particularly close link" between the state of habitual residence and the center of the victim's interests.[58]

[52] Case C-68/93, Fiona Shevill and Others v. Presse Alliance, 1995 E.C.R. I-415.

[53] Case C-509/09, eDate Advertising GmbH and Others, 2011 E.C.R. I-10269. For a discussion see Christian Heinze, *Surf Global, sue local! Der Eurpäische Klägergerichtsstand bei Persönlichkeitsverletzungen im Internet*, *EuZW* 947 (2011); Paul David Mora, *Jurisdiction and Applicable Law for infringements of personality rights committed on the internet*, 34 E.I.P.R. 350 (2012); Matthias Klöpfer, *Persönlichkeitsrechtsverletzungen über das Internet: Internationale Zuständigkeit nach EuGVVO und anwendbares Recht*, JA 165 (2013), and Peter Picht, *Von eDate zu Wintersteiger*, *GRUR Int* 19 (2013). [54] *Id.* § 47. It should be noted that in this case the CJEU clearly refers to the harm suffered "by the holder of a personality right." Such terminology clearly shows that the CJEU intended to limit the effects of the *eDate* judgment only to cases concerning jurisdiction over infringements of personality rights and does not extend to other immaterial rights such as IP rights.

It is clear that the CJEU approach adopted in *eDate* is applicable only to jurisdiction in cases concerning the online infringement of personality rights. With regard to the infringement of IP rights online, two additional cases should be addressed. In *Wintersteiger*,[59] the plaintiff was a proprietor of an Austrian trademark, "Wintersteiger," since 1993.[60] The plaintiff sued a German corporation, Products 4U, for alleged trademark infringement. The German defendant was a manufacturer of various machines used for the service and maintenance of skis and snowboards. Products 4U was also selling online ski accessories produced by Wintersteiger and other manufacturers. Products 4U used the keyword "Wintersteiger" as an AdWord for searches conducted on Google's German web-search site without the permission of Wintersteiger. As a result, if a third person inserted into the search field a keyword "Wintersteiger," alongside natural search results would be sponsored links with the heading "Advertisements" referring to the defendant's ads which did not include a keyword 'Wintersteiger'.

Wintersteiger brought an action before Austrian courts arguing that the placing of an advertisement had infringed its Austrian trademark and sought for an injunction and protective measures. The defendant contested jurisdiction of Austrian courts by arguing that the advertising on google.de is directed exclusively at German customers.[61] The CJEU rendered a two-pronged judgment. First, the CJEU decided that – for the purposes of Article 7(3) – the place of damage is the state where the allegedly infringed trademark is registered. Although the CJEU referred to the 'center of interests' approach put forward in the *eDate* judgment,[62] the Court also made it clear that infringements of personality rights differ from infringements of national registered marks whose legal effects are limited to the territory of the granting/registering state.[63] Second, the CJEU also provided some guidelines for the interpretation of the notion of "the place where the event giving rise to the damage". The CJEU rightly noted that the location of the server is usually uncertain, therefore it can not be used as a foreseeable ground for conferring jurisdiction.[64] Therefore, the definite and identifiable place where the event giving rise to the damage occurs should be the place where the advertiser undertook acts to activate the display on the website; and this place shall be deemed to be the advertiser's place of establishment for the purposes of Article 7(3).

[55] *Id.* § 40.

[56] *Id.* § 48.

[57] *Id.* § 50.

[58] *Id.* § 49.

[59] Case C-523/10, Wintersteiger AG v. Products 4U Sondermaschinen GmbH.

[60] It appears that the same trade mark was also protected in other states, including Germany. *See* Opinion of Advocate General Cruz Villalón, Case C-523/10, Wintersteiger, §§ 4 and 14.

[61] *Supra* note 59, § 13.

[62] *Id.*, §§ 22-23.

In October 2013, the CJEU rendered the most recent judgment in the case *Pieter Pinckney v Mediatech.*[65] In this case, a French author, composer and performer filed a lawsuit in the court of his own residence in Toulouse against the Austrian company seeking compensation for copyright infringement damages. The French plaintiff argued that a CD containing 12 of his copyright-protected songs were produced in Austria without his consent and later offered for sale on various websites of UK intermediaries to whom the right to distribute the CDs was assigned by the Austrian defendant. Those English websites were also available in Toulouse where the author was resident.

The CJEU referred to its previous case-law and reiterated the current state of affairs with regard to application of Article 7(3). The Court first noted that copyrights are similar to industrial property rights because one of the conditions for filing a suit is that the allegedly infringed right is protected in the forum state.[66] In the case of trade marks, protection can be acquired by registering a trade mark. Within the EU, creative works enjoy automatic protection by virtue of Directive 2001/29.[67] Then the Court held that Article 7(3) of the Brussels I Regulation (recast) deals with jurisdiction only; and that the requirement to determine the existence of the harmful event in the forum state should depend on conditions for liability under the potentially applicable substantive laws.[68] In the view of the Court, establishing jurisdiction Article 7(3) does not require allegedly infringing activities to be "directed" to the forum state.[69] However, one of the pre-conditions for bringing a claim is that the IP right is protected in the forum state.[70] Moreover, the court decided that in cases where the allegedly copyright infringing content is reproduced online and could be accessed in the forum state, the court of that state can assert jurisdiction pursuant to Article 7(3). Nevertheless, the CJEU took a strictly territorial view and indicated that the court which exercises jurisdiction pursuant to Article 7(3) can only decide over the copyright infringement in the forum state, because granting jurisdiction to decide over infringing activities in other states would substitute authority of courts in those other states.[71]

[63] *Id*., § 25.

[64] *Id*., § 36.

[65] Case C-170/12, *Pieter Pinckney v Mediatech*.

[66] § 39.

[67] Directive 2001/29/EC of the European Parliament and of the Council of 22 May 2001 on the harmonization of certain aspects of copyright and related rights in the information society, 2001 O.J. (L167) 10.

[68] § 41.

[69] § 42.

[70] § 43.

In January 2015, the CJEU rendered another notable judgment in the case *Hejduk*.[72] In this case a claim was brought by an Austrian professional photographer against a German Corporation which published on its website photographs taken by Ms. Hejduk without her permission. The action was brought before an Austrian court on the ground that copyright infringement occurred in Austria. The Court first of all reiterated its finding in previous case Pinckney and stated that the meaning of the place of the alleged damage varies depending on the nature of the right allegedly infringed.[73] The Court rejected the defendant's argument and followed its earlier case holding that the targeting of allegedly infringing activities is not relevant for assessing the location of the place of damages.[74] It further indicated that accessibility of a website containing infringing photographs in the forum state is sufficient to exercise jurisdiction based on Article 7(3).[75] The Court also touched upon the question regarding the original court's power to assess damages. Yet, this issue was left open: in its judgment the Court only referred to the territoriality principle which was deemed to support the narrow interpretation of court's powers under Article 7(3).[76]

### *(b) Personal Jurisdiction in the US*

The principles of asserting jurisdiction in the United States were developed by state and federal courts. One of the landmark judgments in this context is *International Shoe Co. v Washington*.[77] In this case, the US Supreme Court decided that in personam jurisdiction may be asserted if the defendant had sufficient minimum contacts with the forum and that such exercise of jurisdiction did not offend conventional notions of fair play and substantial justice. [78] The minimum-contacts requirement is met if the defendant purposefully avails himself of the privilege to engage in activities within the forum state, thus invoking the benefits and protection of the laws of that state. Accordingly, a US court can assert personal jurisdiction over a corporation that delivers its products into a stream of commerce with the expectation that they will be purchased by consumers in the forum state.[79]

---

[71] §§ 46-47.

[72] Case C-441/13, Pez Hejduk v. EnergieAgentur.NRW GmbH.

[73] *Id*., § 29.

[74] *Id*., §§ 32-33.

[75] *Id*., § 34.

[76] *Id*., §§ 35-36.

[77] International Shoe Co. v. Washington, 326 U.S. 310, 316 (1945).

[78] *Id*.

US courts have also applied forum non conveniens doctrine in deciding whether personal jurisdiction over foreign defendants could be exercised. Forum non conveniens means that a court that has jurisdiction over a case chooses not to exercise jurisdiction, deferring the case to foreign courts. In deciding to decline jurisdiction, the court must consider two elements: first, there has to be an alternative forum that has jurisdiction to hear the same dispute; and, second, the court has to weigh whether the chosen forum would be more convenient to decide the dispute and where the adjudication of the dispute would best serve the ends of justice.[80] Moreover, in deciding whether it is convenient to decide the case, the court must weigh public and private interests, which include access to proof, availability of witnesses, and all other practical problems which would make the trial of the case easy, expeditious and inexpensive.[81] Yet the fact that foreign law would have to be applied is not sufficient to dismiss a case.[82] In the context of copyright infringements, US courts tend to hold that if an allegedly infringing act occurred abroad and the dispute arose between foreign nationals, there are strong policy concerns to allow dismissal of an action on the grounds of the forum non conveniens doctrine.[83]

In cases concerning infringing acts committed by defendants who are not domiciled in the forum state, US courts could exercise its so-called "long-arm" jurisdiction. For instance, New York courts could exercise jurisdiction over any non-domiciliary who commits a tortious act outside of the state causing injury to person or property within the forum state, if she expects or should reasonably expect the act to have consequences in the forum state and derives substantial revenue from interstate or international commerce.[84] US courts have utilized this long-arm jurisdiction over various commercial cases, including copyright cases. However, in order to successfully proceed to the merits,

[79] World-Wide Volkswagen Corp, v. Woodson, 444 U.S. 286, 297-98 (1980).

[80] Gulf Oil Corp. v. Gilbert, 330 U.S. 501, 508 (1947).

[81] *Id*.

[82] Piper Aircraft Co. v. *Reyno* 454 U.S. 235, 260 (1968).

[83] *See*, *e.g.*, Dominic Murray v. BBC, 81 F.3d 287, 290 (2d Cir. 1996); Skelton Fibres Limited et al v. Antonio Linares Canas et al, 1997 U.S. Dist. LEXIS 2365; *Boosey & Hawkes Music Publishers Ltd v. The Walt Disney Company and Buena Vista Home Video* 145 F.3d 481, 491 (1998).

[84] N.Y. C.P.L.R. 302(a)(3)(ii).

the plaintiff must demonstrate that five criteria for the existence of jurisdiction are met.[85]

In the context of jurisdiction over online copyright infringements, some recent judgments from the Southern District of New York provide insightful considerations. In *Penguin v. American Buddha*, [86] the plaintiff was a famous publisher Penguin Group Inc., incorporated in Delaware, with its principal place of business in New York. The defendant American Buddha was a non-profit corporation with a principal place of business in Arizona. American Buddha maintained a website known as the Ralph Nader Library.[87] The library website was accessible to anyone and provided access mainly to works of classical literature. The information available in the library website was stored on servers located in the states of Arizona and Oregon. As soon as Penguin found out about the library website and that four of its printed works were available there, it filed a lawsuit against American Buddha in the Southern District of New York court claiming that the uploading and making available of the four books infringed Penguin's copyright. American Buddha challenged the jurisdiction of the New York court. The New York District court agreed with the defendant indicating that the place of the injury was 'where the books were electronically copied'.[88] Namely, the Court decided the injury occurred in the place where the books were electronically copied, that is, where the servers were located (Arizona or Oregon) and not in New York, where the Plaintiff Penguin was located, and dismissed the case.

Later the case was brought before the Court of Appeals,[89] which narrowed the question regarding the jurisdiction of New York Courts to the case at hand. It asked, for purposes of determining long-arm jurisdiction, in copyright infringement cases where a copyrighted printed work is uploaded onto the Internet, is the situs of injury the location of the infringing action *or* the residence or location of the principal place of business of the copyright holder?[90] The court held, that in order to establish jurisdiction under New York's long-arm statute covering out-of-state defendant who commits a tortious act abroad, plaintiff is required to show that: (1) defendant's tortious act was committed outside New York; (2) cause of action arose from that act, (3) tortious act caused injury to person or property in New York, (4) defendant expected or should reasonably have expected that his or her action would have consequences in New York, and (5) defendant derives substantial revenue from interstate or international commerce. The Court concluded that, given the circumstances in the Penguin case, the situs of injury was the location of the copyright holder. In the opinion of the Court, the ubiquity of the Internet makes it very difficult to identify and quantify the injury due to the convergence of some of the requirements for long-arm jurisdiction (namely, (a) a tortious act causing injury and (b) the requirement that the non-domiciliary defendant can anticipate where the effects of the act are produced as well as the revenues from interstate and international commerce). The ubiquity of the Internet makes the criterion of "the place where the plaintiff lost business"[91] obsolete because there is no such singular location in the world wide web.[92] In addition, besides the ubiquity of the potentially harmful effects that arise in the case of the online copyright infringements, the Federal Court also highlighted the need to preserve some of the economic incentives for publishers.[93]

---

[85] Namely, that (1) the defendant's tortious act was committed outside New York, (2) the cause of action arose from that act, (3) the tortious act caused an injury to a person or property in New York, (4) the defendant expected or should reasonably have expected that his or her action would have consequences in New York, and (5) the defendant derives substantial revenue from interstate or international commerce.

[86] Penguin Group USA Inc v. American Buddha, 946 N.E.2d 159 (CA N.Y., 2011); Penguin Group USA Inc. v. American Buddha, 640 F.3d 497 (2011); *also* Penguin Group (USA) Inc. v. American Buddha, 2009 US Dist. LEXIS 34032 (S.D.N.Y., 21 April 2009).

[87] *See* naderlibrary.com.

[88] Penguin Group (USA) Inc. v. American Buddha, 609 F.3d (2nd Cir. 2010) ("American Buddha II") at 32.

[89] *Id.*

[90] Penguin Group USA Inc v. American Buddha, 946 N.E. 2d 159, 161 (CA N.Y., 2011).

Another notable copyright infringement case is *Mavrix*,[94] where the plaintiff, Mavrix Photo was a Florida-based corporation which was selling candid celebrity photographs to news portals. Since most of the photographs were usually made in Southern California, Mavrix also had an office in Los Angeles and employed photographers there. The defendant, Brand Technologies, was an Ohio-based corporation which operated a popular Internet site, www.celebrity-gossip.net. This website allowed users to view various media files as well as participate in polls and discussions. In addition, the website also contained a number of ads and links to other sites. The dispute concerned the unauthorized copying and dissemination of pictures of two celebrities taken by a photographer who was working for Mavrix. Mavrix filed the claim in California seeking an injunction against Brand and compensation for actual and statutory damages.

At the first instance, Ohio-based Brand succeeded in challenging the jurisdiction of the Californian court. The Court of Appeals found that the requirements of general jurisdiction were not satisfied; yet, the Californian courts could exercise specific jurisdiction over a copyright infringement action if the elements of a three-prong test are met: (1) The non-resident defendant must purposefully direct his activities to the forum or purposefully avail of the privilege of conducting activities in the forum; (2) the claim must be one which arises out of or relates to the defendant's forum-related activities; and (3) the exercise of jurisdiction must comport with fair play and substantial justice.[95] With regard to the first requirement, the court applied the "effects" test that focuses on the forum in which the defendant's actions were felt, regardless of whether the actions themselves occurred within the forum. The "effects" test analyzes whether the defendant allegedly must have (1) committed an intentional act, (2) expressly aimed at the forum state, (3) causing harm that the defendant knows is likely to be suffered in the forum state.[96]

---

[91] *See* American Eutectic Welding Alloys Sales Co. v. Dytron Alloys Corp., 439 F.2d 428, 433 (2d Cir. 1971).

[92] *Supra* note 88, at 163-164.

[93] *Id.* at 164-164: "[i]f publishers cannot look forward to receiving permission fees, why should they continue publishing marginally profitable books at all? And how will artistic creativity be stimulated if the diminution of economic incentives for publishers to publish academic works means that fewer academic works will be published?"

[94] Mavrix Photo Inc. v. Brand Techs Inc., 647 F.3d 1218 (9th Cir. 2011).

## 4. Is Home Court a Better Fit to Adjudicate Infringements in the Cloud?

The emergence of cloud computing poses a plethora of new legal issues related to the protection of personal data as well as the enforcement of IP rights. Although this technological development has been anticipated, history shows that the legal framework often lags behind swiftly developing social relations. Lawmakers usually make necessary adjustments to the legal framework ex post. Building upon the previous study of the legal framework and recent developments in civil and common law countries, this section aims to stimulate discussion in two ways. First, a brief overview is provided of some of the recent legislative initiatives that sought to propose various adjustments in order to increase the efficiency of cross-border adjudication of IP disputes. Cloud computing technologies and their legal implications were considered by the drafters of those legislative proposals by trying to clarify the notion of the place of tort in the digital environment and the allocation of jurisdiction over such infringements online. The second part of the following section takes a step further and aims to offer some considerations as to whether there should be an additional forum, besides the defendant's residence, where the entirety of the copyright infringement dispute could be adjudicated. We argue that despite courts' attempts to stick to a very narrow approach in exercising jurisdiction over multi-state copyright infringements, there are certain possible types of copyright infringement cases where the plaintiffs (copyright holders) could be allowed to sue at home and seek cross-border relief.

[95] *Id.* at 1127-1228.

[96] *Id.* at 1228.

## 4.1. Recent Legislative Proposals

The discussion concerning jurisdiction over copyright infringement in the cloud would be obsolete without recourse to the recent legislative proposals drafted by several expert groups. These specialized groups emerged when it became clear that the Hague Judgments Project[97] would not achieve its original objectives. Since the harmonization of IP enforcement aspects by an international treaty seems to be rather unlikely in the near future, more comprehensive sets of rules dealing with jurisdiction, choice of law and recognition and enforcement of foreign judgments in disputes concerning the exploitation of IP rights were prepared by experts in the United States,[98] Europe[99] and Asia.[100] These sets of Principles were drafted in light of particular legal traditions and had specific legal goals. For instance, the ALI Principles were drafted with the aim that they could provide some normative guidelines for the adjudication of multi-territorial IP disputes. The CLIP Principles were tailored on the basis of the existing EU practices and with an intention to influence the legal process in Europe. Similarly, one of the objectives of the Transparency Principles was to provide some practical proposals for the Japanese lawmaker that, at the time of drafting, was preparing new rules on international civil litigation. Finally, it appears that the drafters of the Joint Japanese-Korean proposal were expecting to prepare a legislative framework for the adjudication of multi-state disputes in South-East Asian countries.[101]

These legislative proposals brought two separate fields of IP and private international law together and sought to reconcile some of the controversies underlying various aspects of adjudication of multi-state IP disputes. The proposals follow the path-dependency approach in the sense that they are not designing rules from scratch, but rather try to accommodate the existing legal doctrines to the needs of the global economy.[102] As a result, the Principles contain some practical suggestions as to how to deal with the matters pertaining to validity of registered IP rights and offer some suggestions on how to streamline adjudication of multi-state IP disputes by favouring coordination and consolidation.

---

[97] See hcch.net/index_en.php?act=text.display&tid=149.

[98] *See* THE AMERICAN LAW INSTITUTE, *supra* note 33.

[99] EUROPEAN MAX PLANCK GROUP ON CONFLICT OF LAWS IN INTELLECTUAL PROPERTY, CONFLICT OF LAWS IN INTELLECTUAL PROPERTY: THE CLIP PRINCIPLES AND COMMENTARY (2013).

[100] "Transparency Principles" available *at* tomeika.jur.kyushu-u.ac.jp/ip/proposal.htm, *see also* BASEDOW, KONO & METZGER (EDS.), *supra* note 4; the so-called "Joint Japanese-Korean Proposal" is *available at* globalcoe-waseda-law-commerce.org/activity/pdf/28/08.pdf.

[101] For a more detailed comparison, *see* Paulius Jurčys, *International Jurisdiction in Intellectual Property Disputes: CLIP, ALI Principles and other Legislative Proposals in a Comparative Perspective*, 3 JIPITEC 174 (2012).

[102] *Cf*. Basedow, *supra* note 19.

The Principles introduce various solutions with regard to the adjudication of copyright infringement disputes in the digital environment. For instance, § 204(1) of the ALI Principles stipulates that an action can be brought in any state in which the alleged infringer has substantially acted or taken substantial preparatory acts to initiate or further an alleged infringement. The courts of the country where such substantial activities took place would have jurisdiction concerning all injuries arising out of the infringing conduct, regardless of the place where the injuries occur. In addition, as regards Internet-related acts, the plaintiff may also bring an action before the courts of the state to which those infringing activities were directed (§ 204(2) ALI). The Japanese Transparency Principles adopt the so-called market effect test and allow Japanese courts to assert jurisdiction over a dispute concerning a ubiquitous infringement if the effects are maximized in Japan (Art. 105).

Jurisdiction rules in the CLIP Principles to a large degree follow the Brussels I Regulation and provide that, as a general rule, a claim for an IP infringement may be brought to the court of the defendant's residence as well as to the courts of the state where the alleged infringement occurs or may occur.[103] Yet due to the fact that the copyright infringing acts in the digital environment could have spillover effects virtually in any country where the Internet is accessible, the CLIP Principles establish a threshold. Namely, the copyright holder can bring an action before the courts of those states where the alleged infringer committed some acts to initiate or further the infringement or where this activity has been directed to that state. In addition, the CLIP Principles further specify the extent of court powers to adjudicate the cases on the basis of infringement jurisdiction rule (i.e., when the defendant is not resident if the forum state). Article 2:203 of the CLIP Principles introduces some value criteria and provides that a court may also have jurisdiction in respect of infringements that occur or may occur within the territory of any other State, provided that the activities giving rise to the infringement have no substantial effect in the State, or any of the States, where the infringer is habitually resident and (a) substantial activities in furtherance of the infringement in its entirety have been carried out within the territory of the State in which the court is situated, or (b) the harm caused by the infringement in the State where the court is situated is substantial in relation to the infringement in its entirety.[104]

These three sets of principles were drafted by groups of practicing lawyers and academics and have no binding force. Nevertheless, they lay a solid foundation for further development of this increasingly significant area of law. The normative suggestions provided in those sets of principles could be materialized by the courts, which could take them into consideration when interpreting and applying national jurisdiction and choice of law provisions in actual cases.

[103] Art. 2:201 of the CLIP Principles.

[104] Art. 2:203(2) of the CLIP Principles.

As stated above, the primary task of the drafters of the legislative proposals was to clarify, to the extent possible, where claims for ubiquitous IP-right infringements could be adjudicated. One of the common features of all of the proposals was the aspiration to curtail the number of available fora for the adjudication of such disputes. It is also apparent that there are various ways to achieve that. For instance, it could be possible to focus on the place where the substantial injury occurs (the ALI approach), or the place where the significant market interests of the right holder are affected (the Transparency Principles approach). It may also make sense, depending on the factual circumstances of the case, to attach jurisdiction to the courts of the place where the infringement was initiated or facilitated (the CLIP approach). Given the diversity of possible kinds of infringements in the digital environment, it would be unreasonable to completely rule out any of those approaches. On the contrary, the ubiquitous nature of tortious acts mandate a more flexible approach and leave some degree of discretion for the court to determine whether jurisdiction should be asserted or not.

## 4. 2. Concentrating Litigation at the Right-Holder's Center of Economic Interests

The universal protection afforded by the Berne Convention to the creators of original content allows the copyright-holder to seek protection in every country where the alleged infringement occurs. From a costs and benefits point of view, it may be questioned whether it is efficient to have a litigation model where a copyright-holder who seeks to protect his rights internationally is compelled to bring infringement actions in every state where the content is available. Such a mosaic framework of litigation exists in many, if not most, legal systems. For instance, Brussels I Regulation (Recast) gives the plaintiff in a ubiquitous infringement suit only one option: to institute proceedings before the defendant's forum under Article 4 and try to recover damages covered in all states for which protection is sought. Otherwise, the copyright-holder has an option to file a suit in every state pursuant to Article 7(3) of the Brussels I Regulation (Recast). A similarly restrictive situation exists also in common law states, where in personam and subject-matter jurisdiction requirements eventually lead to concentration of litigation in the defendant's forum with very limited possibilities of cross-border legal redress.

In ubiquitous copyright infringement cases there are several patterns of litigation. In most cases, the copyright holder sues an infringer seeking some sort of legal redress (injunctive relief or compensation for damages). If we look more accurately, plaintiffs in such cases could be either major corporations (preeminent publishers, record labels, software developers) or individual authors. Moreover, regardless of their economic power, both large copyright holders as well as individual creators prefer to file copyright infringement actions in their home state. Such a possibility of filing an action at the copyright owner's home court, while controversial at first sight, deserves further analysis. The realities of digital communication and ubiquity of copyright infringements online have completely changed the territorial nature of IP litigation. Hence, it is necessary to examine in what situations the accessibility of a copyright infringing material in the plaintiff's forum state could be a sufficient ground to establish jurisdiction over a non-resident defendant and what types of copyright infringement disputes could be adjudicated before the plaintiff's forum. It is also necessary to examine in what cases it would be more efficient to allow the court at the plaintiff's (copyright-holder's) home court or court of plaintiff's centre of interests to exercise jurisdiction over copyright infringements abroad.

One may argue that the court of a state where the infringing content is available may also be considered as having sufficient personal jurisdiction over a non-resident defendant. This could occur in cases where substantial effects of the defendant's activities occur in the state where the copyright holder has his residence or centre of economic interests. Arguably, opening cross-border proceedings against a non-resident defendant could prevent excessively costly litigation in remote states where neither of the parties has any personal points of attachment. Territorial balkanization of adjudicatory powers is neither efficient in dealing with the activities in globally interconnected markets, nor does it add any certainty or predictability to the parties involved.

Four main considerations are noteworthy in this discussion. First, such mosaic adjudication of ubiquitous copyright infringement disputes does not often occur in practice; in most cases the plaintiff chooses one or several states to resolve the dispute. In addition, previous discussion shows that in many cases plaintiffs try to seek judicial redress before their home courts. If the plaintiff is able to begin court proceedings at his/her home court on the basis that this is the place where the damage was sustained, why should it not be possible to adjudicate the entirety of the dispute? Concentration of ubiquitous copyright infringement disputes before courts of the state where the copyright holder has his/her center of economic interests seems to offer more legal certainty and predictability to plaintiffs, defendants, and courts. Defendants are presumably more likely to be aware where the center of plaintiff's economic interests is, and could anticipate possible adjudication of the entire dispute in that state. Hence, differently from dissemination of printed material, the reality of ubiquitous infringements online offers additional justifications to concentrate proceedings before the courts of the state where the right-holder's center of economic interests are located. This approach could be also justified because it would preempt possible parallel proceedings on an international level.

The second issue relates to the taxonomy of copyright infringement disputes. More specifically, it is necessary to carefully consider whether it would be reasonable to allow the right holder to bring *any* kind of copyright infringement dispute before the home court or the court of the place where the right-holder has his/her centre of economic interests. Proponents of the most liberal view could argue that copyright holders should be allowed to bring any type of infringement action before the court of a state where the plaintiff has his/her center of economic interests, i.e. infringement of any kind of economic or moral right. The critics of such a liberal approach are those who see great advantages of strict territoriality in copyright law. The advocates of a strictly territorial approach could argue that the scope of protection of an author's economic rights (e.g., right of reproduction, or right of distribution) is strongly connected to the protecting state's economic policy; and that only the courts of the protecting state should be competent to adjudicate such cases. Certainly, strict territoriality and mosaic adjudication of ubiquitous copyright infringement disputes offers much legal certainty. Yet the question is whether legal certainty should be deemed as a greater good. In complex ubiquitous infringement cases, parties' welfare could be best measured in the efficiency of adjudication which ideally would mean concentration of the proceedings in one forum and prevention of parallel litigation. Litigation framework existing pursuant to Articles 4(1) and 7(3) of the Brussels I Regulation (Recast) could be viewed as an attempt to balance legal certainty and efficiency. However, recently emerging disputes over ubiquitous copyright infringements show that it is necessary to look for more efficient solutions for those cases where the forum country and protecting countries overlap, but not necessarily coincide.

Much of the earlier scholarly discussion focused on the question whether any parallels could be drawn between infringements of personality rights and copyrights. Scholars and courts stumbled upon the question whether similar standards to assert jurisdiction could be applied in cases concerning infringements of personality rights and infringements of author's moral rights. The intersection between the personality rights and copyright is apparent especially if one sees copyright as akin to the personality of an author (*droit d'auteur*).[105] Yet, rather than tottering around this idle relationship dilemma, conflict of laws scholars should give a second thought to the question whether it is reasonable to allow the courts of the state in which the copyright holders interests are centered to exercise cross-border jurisdiction over claims related to infringement of author's moral rights. The personality theory of copyright emphasizes the strong connection between an author and his/her work and could therefore be a strong argument to support the view that a court should be allowed to assert jurisdiction over claims concerning infringements of author's moral rights in the forum state as well as in foreign states. It is true, that authors' moral rights may not be protected in all countries of the world, but this issue should be considered during the substantive-law analysis after the court decides to exercise cross-border jurisdiction.[106]

[105] Justin Hughes, *The Philosophy of Intellectual Property* 77 GEORGETOWN LAW JOURNAL 287 (1988) and ROBERT P. MERGES, JUSTIFYING INTELLECTUAL PROPERTY (2011); Cyrill P. Rigamonti, *Deconstructing Moral Rights*, 47 Harv. Int'l L.J. 353 (2006).

The third controversial question is under what conditions a court of the state where the copyright holder has his/her center of economic interests could exercise powers to order provisional and protective measures with cross-border effects and decide over damage sustained in other states. Is mere evidence of the plaintiff's center of economic interests in the forum state sufficient for exercising cross-border jurisdiction and deciding over copyright infringements in foreign states? In order to trigger the discussion of this question, we suggest examining the implications of the "market effects" test in adjudicating ubiquitous copyright infringements. The market-effects test could be seen as an umbrella concept encompassing multiple factual considerations (accessibility, language, domain name, number of hits or the location of the server/PC as well as effects on the potential market of the allegedly infringed work). In many ubiquitous copyright infringement cases, the place where the right-holder has his/her center of interests overlaps with the market where the defendant's activities have effects. Center of interests and market effects as jurisdiction criteria appear to be flexible in localizing ubiquitous copyright infringements and deciding whether a court of a particular state should assert jurisdiction over such disputes.

The market-effects test and the determination whether the copyright holder's center of affected interests are in the state whose courts are seized could play an instrumental role in establishing long-arm jurisdiction over non-resident defendants. The US could provide some helpful guidance in this regard: one of the conditions for exercising long-arm jurisdiction is finding that a foreign defendant reaps some economic benefits by making the copyright infringing content available in the forum state.[107]

The market-effects test could help to avoid the mosaic application of multiple laws which stem from the lex loci protectionis principle by allowing the court to apply the laws of one or several states whose markets are most affected by a ubiquitous copyright infringement.[108]

---

[106] *Cf.* Pinckney, supra note 65; § 42 where the CJEU indicated that the place of harmful event for jurisdiction purposes "can not depend on the criteria which are specific to the examination of the substance" of the claim.

[107] *See* e.g., International Shoe Co. v. Washington, 326 U.S. 310 (1945) and World-Wide Volkswagen Corp. v. Woodson, 444 U.S. 286 (1980).

[108] See Art. 8 of the Rome II Regulation; Ryu Kojima, Ryo Shimanami & Mari Nagata, *Applicable Law to Exploitation of Intellectual Property Rights in the Transparency Proposal, in* BASEDOW, KONO & METZGER (eds.), *supra* note 4, at 181-200; Paulius Jurčys, *Applicable Law to Intellectual Property Infringements in Japan: Alternatives to the Lex Loci Protectionis Principle*, 24 INT'L REV. L. COMP. &

It should also be noted that in many multi-state IP infringement cases plaintiffs try to bring claims for infringement of IP rights as well as unfair competition claims. In such cases, the application of a single connecting factor to determine the law governing different claims could streamline the adjudication process. Moreover, the application of the market-effects test could also function as a de minimis threshold helping to avoid adjudication of ubiquitous copyright infringement disputes and/or applying laws of the states that bear little connection to the actual infringement.

Fourth, cross-border adjudication of ubiquitous copyright infringement disputes before the courts of the right holder's center of economic interests should be a balanced test that takes into account economic power of all parties to a dispute. Certainly, such a concentrated adjudication regime before the plaintiff's forum might enormously strengthen the position of copyright holders. This could be problematic, especially when the plaintiff in a copyright infringement dispute is an economically powerful corporation which is, in any case, capable of seeking protection of its rights abroad. So the real question mainly deals with the adjudication of claims brought by individual copyright holders who do not possess sufficient resources for filing actions in each and every state in which they would like to protect their rights. Due to excessive costs of litigation, these individual authors or performers decide to file claims before their home courts or courts of the state where their economic interests are centered. Hence, there is a strong policy reason to establish additional safeguards for economically weaker creators such as freelance photographers, authors or performers. Similar protections are already entrenched in jurisdiction statutes of civil law countries for cases involving individual consumers or employees.[109] Hence, center of interests should be conceptualized as a case-specific criterion that takes into consideration various factual circumstances of the dispute. The adjudication of the entirety of a ubiquitous copyright infringement dispute before the courts of the state where the copyright holder has his/her economic interests should be possible if the court believes that there is a reasonable balance between the interests of the parties and that such an adjudication provides for an equitable solution of the case.

## 5. Concluding Remarks

The Internet opened the gates to a vast pool of knowledge, available at very low cost and in multiple forms. Cloud computing is another unprecedented technological development. The possibility of storing and accessing big data in a remote server provides new business

---

TECH. 193 (2010); Carl F. Nordmeier, *Cloud Computing und Internationales Privatrecht – Anwendbares Recht bei der Schädigung von in Datenwolken gespeicherten Daten*, MMR 151 (2010).

[109] *See*, *e.g*., Arts. 17 and 20 of the Brussels I Regulation (Recast).

opportunities and even further reduces the costs of communication. The ubiquitous flow of information in the cloud-based environment has brought about a number of questions concerning traditional legal approaches related to the allocation of property rights, data protection, and conflict of laws. The statement of Pirate Bay at the beginning of this article perfectly illustrates the difficulties arising in relation to the control of communication in the cloud. Grid cloud computing further contributed to the irrelevance of the location of servers for the establishment of territorial connecting factors. Cloud computing disrupted traditional conflict of laws methodology which rests upon Savigny's idea that each legal relationship has a "seat" in a particular state. Although a proper international litigation system must be founded on a connection between the dispute and a forum state, existing principles of exercising jurisdiction deserve a close reconsideration.

This paper aimed to highlight the main rules and principles that have been employed by courts in civil and common law countries in dealing with cross-border copyright infringement disputes. The territorial fragmentation of IP rights and resulting state-by-state adjudication of IP disputes is neither efficient nor does it fit the needs of the digital economy.

In particular, this paper touched upon the question whether it is reasonable to allow adjudication of territorially unlimited claims for ubiquitous copyright infringement before the courts of the state where the right holder has his/her center of economic interests. This article predicts that the main legal problem will be the establishment of sufficient personal jurisdiction over a non-resident defendant. This could be achieved in cases where the activities of the defendant show that there is a significant market effect in the state where the copyright holder has his/her center of economic interests. For instance, one such case could arise when the defendant may purposefully avail himself by targeting a specific country or by making profits from the dissemination of the infringing content in that state. However, it has been also shown in this paper that in recent decisions courts started to realise that "targeting" requirements should not be necessary to establish the place of actual damage. In this paper we suggest that the existing copyright litigation regime could be relaxed especially in cases where economically weaker copyright holders (especially, authors, performers) seek legal redress before their home courts (courts where they have their center of economic interests) for the damage caused by ubiquitous infringements. This is not a new idea, for similar plaintiff-favoring jurisdiction rules exist in cases involving consumers or employees.[110] We also argue that from a cost and benefit perspective, allowing the court of the place where the copyright holder has his/her economic interests to assert jurisdiction over ubiquitous infringements could create an incentive for parties to reach settlement agreements before a court resolves the dispute. We conclude that the plaintiff's center of economic interests is a powerful and thought-provoking theory which deserves much more attention in the future and could help develop a more case-specific framework for adjudication of ubiquitous copyright infringement cases.

[110] *See* Arts. 17 and 20 of the Brussels I Regulation (Recast).

This whole set of problems has been carefully analysed by a special committee to deal with the intersection of IP and private international law, created in 2010 under the auspices of the International Law Association (ILA).[111] Some members of this ILA Committee have played a leading role in drafting the aforementioned Principles. Other members were experts from regions previously not represented in the drafting of the Principles. The Committee conducted a comparative study of the legislative proposals[112] and published two reports where various issues related to adjudication of ubiquitous copyright infringement cases were addressed.[113] One of the main objectives of the Committee was to prepare a Resolution that would contain guidelines for international organizations, such as the Hague Conference on Private International Law and WIPO, as well as regional and national lawmakers. The members of the Committee are also investigating possible ramifications of cloud computing to the adjudication of multi-state copyright-related disputes and expect to come up with some viable proposals in the near future.

[111] ila-hq.org/en/committees/index.cfm/cid/1037.

[112] Comparative reports are published in JIPITEC, see *supra* note 4.

[113] *See* law.kyushu-u.ac.jp/programsinenglish/ila2012/index.htm.